\documentclass[12pt]{article}
\usepackage{amsmath,amssymb}

\usepackage{graphicx}
\usepackage{wrapfig}

\usepackage{hyperref}
\usepackage{cite}

\def\beq{\begin{eqnarray}}
\def\eeq{\end{eqnarray}}

\def\o{\over}

\newcommand{\Tr}{\,\mathrm{Tr}\,}            

\newcommand{\be}{\begin{equation}}
\newcommand{\ee}{\end{equation}}
\newcommand{\bea}{\begin{eqnarray}}
\newcommand{\eea}{\end{eqnarray}}
\newcommand{\bg}{\begin{gather}}
\newcommand{\eg}{\end{gather}}
\newcommand{\bseq}{\begin{subequations}}
\newcommand{\eseq}{\end{subequations}}

\renewcommand{\ln}{\mathop{\rm ln}\nolimits}

\def\tr{\hbox{Tr}}

\def\be{\begin{eqnarray}}
\def\ee{\end{eqnarray}}
\def\lb{\label}

\begin{document}

\title{\textbf{Newton constant, contact terms and entropy }}

\vspace{1cm}
\author{ \textbf{Sergey N. Solodukhin  }} 

\date{}
\maketitle

\begin{center}
  \hspace{-0mm}
  \emph{Laboratoire de Math\'ematiques et Physique Th\'eorique CNRS-UMR
7350}\\
  \emph{Universit\'e Fran\c cois-Rabelais Tours F\'ed\'eration Denis Poisson - CNRS, }\\
  \emph{Parc de Grandmont, 37200 Tours, France} \\    
\end{center}

{\vspace{-11cm}
\begin{flushright}
 NSF-KITP-12-167
\end{flushright}
\vspace{11cm}
}



\begin{abstract}
\noindent {    We discuss the renormalization of the Newton constant due to fields of various spin $s$. We first briefly review the cases
of $s=0, \, 1/2, \, 1,\, 3/2$ already discussed in the literature and notice the appearance of the well-known contact terms for the vector bosons. We then extend 
this discussion of the contact terms to massive vector fields, $p$-forms and to the case 
of spin $s=2$ particles (gravitons).  We observe that, in general,  the contact terms originate from the fields which mediate the interactions (such as vector gauge bosons and gravitons).
We then discuss entanglement entropy and the conical entropy and their relation to the renormalized Newton constant. 
We address the puzzle  of the non-analytic terms due to fields of spin $s=2$ and suggest that the resolution of this puzzle comes from the non-equivalence of the orbifold and $n$-fold cover constructions which are used in the entropy calculations.
Finally, we propose a mechanism
by which the Bekenstein-Hawking entropy is identified with entanglement entropy  in any theory which includes both matter fields and the mediators of interactions (vector gauge bosons and gravitons).   }
\end{abstract}

\vskip 2 cm
\noindent

\newpage
    \tableofcontents
\pagebreak

\newpage

\section{Introduction}
\setcounter{equation}0
The Newton constant not only defines the strength of gravitational interaction in a low energy regime
but also determines the value of entropy associated with the black hole horizons \cite{BH}. Both aspects arise
in a semi-classical approximation in which the graviton loops \cite{'tHooft:1974bx} do not play an important role and, as it seems, can be ignored.
The statistical aspect of horizons,  however, is believed to be a window through which one may try to see 
the fundamental degrees of freedom of Quantum Gravity otherwise unaccessible. In the recent years there has been proposed a number of approaches
that explore this window in various directions \cite{Strominger:1996sh}, \cite{HCFT}, \cite{Dvali:2011aa}.  On the other hand, 
it seems plausible that, in a quantum theory,  the observable  Newton constant and the entropy should arise  simultaneously 
within the same mechanism which deals with the observable low energy particles, irrespectively to the details of the hypothetical   more fundamental description.

The resolution of the problem in this  formulation  boils down to the analysis of the renormalization of the Newton constant and the appropriate calculation of the entropy \cite{Susskind:1994sm}. The entropy most suitable to this purpose is the entanglement entropy \cite{EE} defined for a given surface as entropy of ignorance of the modes residing 
inside the surface, for the earlier works in this direction see \cite{S} and more recently \cite{DS}, \cite{M}, \cite{L}, \cite{Jacobson:2012ek}. This definition is very much suitable to describe the situation with the principle impossibility to gain information on that physical modes
which propagate inside a black hole (for a review see \cite{Solodukhin:2011gn}, \cite{Jacobson:2012yt}).  The departure point in this direction is the observation that the entanglement entropy, if combined with   the classical Bekenstein-Hawking entropy, can be expressed in terms of the  renormalized  Newton constant \cite{Susskind:1994sm}, \cite{Solodukhin:1994yz}, \cite{ted},  \cite{Fursaev:1994ea}, \cite{Demers:1995dq}. If it worked for all fields this would allow us to relate directly the two aspects, the generation of the observable Newton constant and the  statistically meaningful Bekenstein-Hawking entropy. It was observed, however, very soon that this does not work for all fields. Namely, it does not work for 
a scalar field non-minimally coupled to gravity \cite{Solodukhin:1995ak} and for the gauge vector fields  \cite{Kabat:1995eq}. Kabat has observed that
the difference is due to the presence of the so-called contact terms which do not appear to have a statistical meaning. The issue of these terms has been actively discussed in
the literature \cite{contact}. These terms are often  considered as obstruction to the successful realization of the program of identifying the Bekenstein-Hawking entropy as entanglement entropy.  

In the present paper we extend the discussion of the contact terms to other fields: massive vector fields, $p$-forms and particles of spin $s=2$ (gravitons). The result for the contact terms due to gravitons 
was already given (without details)  in \cite{Solodukhin:2011gn}. Here we give a detail analysis of the linear perturbations in the Einstein-Hilbert action in arbitrary dimension
$d\geq 4$. We decompose the gravitational perturbations of the Einstein-Hilbert  action on the trace (conformal mode) and the traceless part and  show that in generic dimension $d>4$ the conformal mode becomes non-minimally coupled so that the coupling vanishes only if $d=4$. Moreover, we show that the relevant operator which governs the
traceless perturbations  is a curvature-dependent modification of the Lichnerowich operator.  The complete treatment of all perturbations allows us to determine the contact terms
due to gravitons.  In general, we notice that the presence of the contact terms is due to the fields which mediate the interactions and thus are due to the presence of the 
(exact or broken) gauge symmetry.  

We then suggest that, considering all fields that appear in a low energy regime, there exists a particular choice of the otherwise arbitrary  bare gravitational  constant that
the Bekenstein-Hawking entropy, defined in terms of the renormalized Newton constant, is identical to the entanglement entropy. The UV cut-off needed for the proper definition of the entanglement across the horizon   then is automatically set  by the Planckian scale. In this mechanism the two pieces in the Bekenstein-Hawking entropy which do not have a statistical meaning, the
one which is defined by the bare Newton constant and the other  due to the presence of the contact terms in the renormalized Newton constant, mutually cancel
so that what is left is genuine entanglement entropy.

\section{Newton constant in quantum action}
\setcounter{equation}0

We start with reminding a general set up for the calculation of the Newton constant which appears in the quantum effective action.
The effective action of a field of spin $s$ can be written as  
\be
W_{(s)}={(-)^{2s}\o 2}\int^\infty_{\epsilon^2}{d\tau\o \tau}\tr e^{-\tau\Delta^{(s)}}\, .
\lb{effective action spin s}
\ee
The second order covariant operators acting on  the spin-$s$ field can be represented in the following general form
\be
\Delta^{(s)}=-\nabla^2\delta_{AB} +X^{(s)}_{AB}\, ,
\lb{diff-operator}
\ee
where  the matrix $X_{(s)}$ depends on the chosen representation of the quantum field and it is linear in the Riemann tensor.
The UV divergences of the action (\ref{effective action spin s}) are analyzed using the small proper time $\tau$  for the heat kernel of  this operator 
\begin{eqnarray}
\tr e^{-s\Delta^{(s)}}(\tau)={1 \over (4\pi \tau)^{d \over
2}}
\sum_{n=0}^{}{} {a}_n \tau^n\, .
\label{K expansion}
\end{eqnarray}
The induced Newton constant is determined by the coefficient $a_1$ in this expansion. It  has the general form
\be
a_1^{(s)}=\int d^dx\sqrt{g}({{D}_s(d)\o 6}R-\tr X^{(s)})\, ,
\lb{a2-general spin}
\ee
where $D_s(d)=\delta_{AA}$ is the dimension of the representation of spin $s$. For $s\leq 2$ spins  we have
\be
&&D_{s=0}(d)=1\, , \,\, D_{s=1/2}(d)=2^{[d/2]}\, ,\,\, D_{s=1}(d)=d\, , \nonumber\\
&&D_{s=3/2}(d)= d\, 2^{[d/2]}\, , \,\, D_{s=2}(d)={(d-1)(d+2)\over 2}\, .
\lb{dimensions representation}
\ee
$D_s(d)$ can be interpreted as the number of off-shell degrees of freedom, for $s=2$ it counts number of degrees of freedom of a symmetric traceless matrix.

The effective actions is expanded in powers of the  curvature. The first term in this expansion is
\be
W^{(1)}_{(s)}=-\frac{1}{16\pi G(\epsilon)}\int d^dx \sqrt{g}R=-{1\o (4\pi)^{d \o 2}(d-2)} {(-1)^{2s}\o \epsilon^{d-2}} a_1^{(s)}
\lb{W}
\ee
where $G(\epsilon)$ is the induced Newton constant, $\epsilon$ is the UV cut-off in the theory.

In what follows we shall give the calculation of the induced Newton constant for fields of various spin $s$. 
The result can be represented in the following form
\be
{1\o 4G(\epsilon)}={1\o (4\pi)^{d-2 \o 2}(d-2)}  {1\o \epsilon^{d-2}}\left({{\cal D}_{s}(d)\o 6}-c_{(s)}(d)\right)\, ,
\lb{Gren-spin-s}
\ee
where ${\cal D}_{s}(d)$ is the number of on-shell degrees of freedom, each fermionic degree of freedom is counted with the weight $1/2$, 
\be
&&{\cal D}_{s=0}(d)=1\, , \,\, {\cal D}_{s=1/2}(d)=\frac{2^{[d/2]}}{2}\, , \,\, {\cal D}_{s=1}(d)=d-2\nonumber \\
&&{\cal D}_{p}(d)=C^p_{d-2}\, , \,\, {\cal D}_{s=3/2}(d)= (d-3)\, \frac{2^{[d/2]}}{2}\, , \,\, {\cal D}_{s=2}(d)={d(d-3)\over 2}\, .
\lb{D}
\ee
where $p$ stands for $p$-forms.
The quantity $c_{(s)}(d)$ which appears in (\ref{Gren-spin-s}) is what we shall call the contact term for field of spin $s$. This quantity is non-zero only for the bosonic gauge
fields: $s=1$ vector bosons, $s=2$ gravitons, $p$-forms,
\be
c^{(s)}(d)=\xi\, , \,\, c^{(s=1)}(d)=1\, , \,\, c^{(p)}(d)=C^{p-1}_{d-4}\, ,\, \, c^{(s=2)}(d)=\frac{d^2-d+4}{2}\, ,
\lb{cc}
\ee
where $\xi$ is non-minimal coupling of scalar field to gravity.
 The cases $s=0,\, 1/2,\, 1$ were already considered in the literature. We shall give below the detailed analysis of the
case of $s=2$ spin and of the $p-$forms.  

We start with some well-known examples (see \cite{Kabat:1995eq}, \cite{Larsen:1994yt}, \cite{LW}, \cite{Fursaev:1996uz}).

\section{Fields of  spin $s<2$}
\setcounter{equation}0

\subsection{ Non-minimally coupled scalar field, $s=0$}
A non-minimally coupled scalar field is described by the operator
\be
\Delta^{(s=0)}=-\nabla^2+\xi R\, ,
\lb{s=0}
\ee
where $\xi$ is a parameter, $\xi=\xi(d)=\frac{(d-2)}{4(d-1)}$ corresponds to the conformal case. The contact term in this case is
\be
c_{(s=0)}(d)=\xi\, .
\lb{cs0}
\ee

\medskip

\subsection{ Dirac fermion field, $s=1/2$} The partition function  of the Dirac fermion, $Z_{1/2}=\det^{1/2}\Delta^{(s=1/2)}$ is expressed in terms of the  squared Dirac operator.
The latter  takes the form (\ref{diff-operator}) with
matrix
\be
X^{(s=1/2)}_{AB}=\frac{1}{4}R \delta_{AB}\, ,
\lb{s=1/2}
\ee
where $R$ is the Ricci scalar and $(A, B)$ are the spinor indexes. Thus we have that
\be
(-1)^{2s}a_1^{(s=1/2)}=-D_{s=1/2}(d)(\frac{1}{6}-\frac{1}{4})\int R=\frac{{\cal D}_{s=1/2}(d)}{6}\int R\, .
\lb{s=1/2}
\ee
The comparison with (\ref{Gren-spin-s}) shows that the contact term vanishes in this case
\be
c_{(s=1/2)}(d)=0\, .
\lb{csf}
\ee

\medskip

\subsection{ Gravitino, $s=3/2$} The partition function, including  the Faddeev-Popov   and the Kallosh-Nelson ghosts contributions, is $Z_{3/2}=\det ^{1/2}\Delta^{(3/2)}\det^{-3/2}\Delta^{(1/2)}$, see for instance \cite{Duff:2002sm}. The operator $\Delta^{(3/2)}$ takes the form (\ref{diff-operator}) with
\be
&&X^{(3/2)}_{AB, \mu\nu}=\frac{1}{4}R\delta_{AB}g_{\mu\nu}-\frac{1}{2}R_{\mu\nu\alpha\beta}(\gamma^\alpha\gamma^\beta)_{AB}\, ,\nonumber \\
&&\Tr X^{(3/2)}=\frac{1}{4}D_{3/2}(d)\int R\, .
\lb{s3/2}
\ee
The  heat kernel coefficient of this operator reads
\be
a^{3/2}_1=D_{3/2}(d)(\frac{1}{6}-\frac{1}{4})\int R=-\frac{1}{12}{D}_{3/2}(d)\int R\, 
\lb{sss}
\ee
so that, taking into account (\ref{s=1/2}), we find that the appropriate heat kernel coefficient in the partition function is
\be
-a_1=-a^{3/2}_1+3a^{1/2}_1=\frac{1}{6}{\cal D}_{3/2}(d)\int R\, ,
\lb{as3/2}
\ee
where ${\cal D}_{s=3/2}(d)= (d-3)\, \frac{2^{[d/2]}}{2}$ is the number of on-shell degrees of freedom for spin $s=3/2$.
Notice that it vanishes in $d=3$ as it should be by supersymmetry\footnote{We thank M. Duff for this remark.}.
The comparison with (\ref{Gren-spin-s}) shows that the contact term vanishes in this case too.

\medskip

\subsection{ Gauge field, $s=1$} The case of gauge vector bosons has been actively considered in the literature. Here we briefly summarize the analysis for the Newton constant.
The partition function of a gauge boson is 
\be
Z_{(1)}={\det}^{-1/2}\Delta^{(1)}_+\det\Delta^{(0)}\, ,
\lb{Z1}
\ee
 where $\Delta^{(1)}$ is the operator of the form  (\ref{diff-operator}) with
\be
X^{(1)}_{\mu\nu}=+R_{\mu\nu}\, , \,\,\, \tr X^{(1)}=R\, .
\lb{s1}
\ee
The heat kernel coefficient of this operator is
\be
a_1^{(1)}=(\frac{d}{6}-1)\int R\, .
\lb{ss1}
\ee
The ghost operator  in (\ref{Z1}) is the minimal scalar $(s=0)$ operator.  Thus, taking into account the ghost contribution one has that
\be
a_1=a_1^{(1)}-2a_0^{(s=0)}=(\frac{d}{6}-1-\frac{2}{6})\int R=(\frac{1}{6}{\cal D}_{s=1}(d)-1)\int R\, .
\lb{a1}
\ee
The contact term thus is
\be
c_{(s=1)}(d)=1\, .
\lb{cs1}
\ee
This is the famous Kabat's contact term  \cite{Kabat:1995eq} (see also \cite{LW}).
We see that the induced Newton constant due to the spin $s=1$ field  is not a priori positive. Indeed, it is negative if $d<8$, vanishing if $d=8$ and positive if $d>8$. 

\subsection{Massive fields}

In the case of scalars or Dirac fermions it is rather straightforward to add mass  into consideration. One simply adds a mass term to  the second order operator
$\Delta_{s,m}=\Delta_s+m^2$, $s=0,\, 1/2 $. Respectively, one modifies the heat kernel,
\be
\Tr e^{-\tau \Delta_{s,m}}=\tr e^{-\tau \Delta_s}\, e^{-\tau m^2}\, ,
\lb{3.1}
\ee
where $\tr e^{-\tau \Delta_s}$ has the decomposition (\ref{K expansion}).
The induced Newton constant then is given by the integral
\be
{1\o 4G(\epsilon)}=\frac{{\cal D}_s}{12}\int_{\epsilon^2}^\infty\frac{d\tau}{(4\pi\tau)^{d/2}}e^{-m^2\tau}\, , s=0,\, 1/2
\lb{3.2}
\ee
To leading order in $\epsilon$ this gives us the same behavior as before for the massless fields. On the other hand, there appear some sub-leading terms
as function of $m^2\epsilon^2$.  

The analysis is more subtle for a massive vector field.  The reason is that the gauge invariance is broken by the presence of the mass term that effectively modifies the number of
the physical degrees of freedom. The analysis made in \cite{Barvinsky:1985an} shows that the partition function of a massive vector field is
\be
Z_{s=1,m}={\det}^{-1/2}(\Delta^{(1)}_++m^2)\, {\det}^{1/2}(\Delta^{(0)}+m^2)\, .
\lb{3.3}
\ee
The respective contribution to the Newton constant is
\be
{1\o 4G(\epsilon)}=\frac{1}{2}\left(\frac{{\cal D}_{s=1,m}}{6}-c^{(s=1,m)}\right)\int_{\epsilon^2}^\infty\frac{d\tau}{(4\pi\tau)^{d/2}}e^{-m^2\tau}\, ,
\lb{3.4}
\ee
where 
\be
{\cal D}_{s=1.m}(d)=d-1\, , \, \, c^{(s=1,m)}(d)=1\, .
\lb{3.5}
\ee
We see that the structure of the induced Newton constant remains the same as in the massless case. The physical degrees of freedom (their number  is  modified if compared to the massless case) contribute positively with the weight $1/6$ while the contact term contributes negatively to the gravitational constant.  Notice, that the contact term in (\ref{3.5}) is the same as in the massless case.

Generalizing (\ref{3.4}) to fields of various spin $s$ and mass $m_s$ we have 
\be
{1\o 4G(\epsilon)}=\sum_{s}\frac{1}{2}\left(\frac{{\cal D}_{s,m}}{6}-c^{(s,m)}\right)\int_{\epsilon^2}^\infty\frac{d\tau}{(4\pi\tau)^{d/2}}e^{-m_s^2\tau}\, 
\lb{3.6}
\ee
for the complete contribution to the Newton constant.

\section{Gauge $p$-forms}
\setcounter{equation}0

The other example of lower spin fields with gauge symmetry are the gauge $p$-forms,
\be
\omega^{(p)}=\frac{1}{p!}\omega_{i_1..i_p}dx^{i_1}\wedge ..\wedge dx^{i_p}\, ,
\lb{omega}
\ee
where $\omega_{i_1..i_p}$ is a totally antisymmetric tensor of rank $p$. The action 
\be
W^{(p)}=\frac{1}{2}\int \star d\omega^{(p)}\wedge d\omega^{(p)}
\lb{Wp}
\ee
is invariant under the gauge transformations,
\be
\omega^{(p)}\rightarrow \omega^{(p)}+d\alpha^{(p-1)}\, ,
\lb{gauge}
\ee
where $\alpha^{(p-1)}$ is a $(p-1)$-differential form. The $p$-form is thus a generalization of the Maxwell gauge field  considered in the previous section.
The fixing of the gauge symmetry (\ref{gauge}) requires introduction of a set of the Faddeev-Popov ghosts, the differential forms of rank $p-k$, $k=1,2,..$ 
some of them considered as bosonic fields and the others as fields of fermionic statistics. The resultant partition function then takes the form (see for instance \cite{Obukhov:1982dt})
\be
Z^{(p)}=\prod_{k=0}^p(\det\Delta_{p-k})^{(-1)^{k+1}\frac{(k+1)}{2}}\, ,
\lb{Zp}
\ee
where $\Delta_k=d^{k-1}\delta^{k}+\delta^{k+1} d^{k}$ is the Beltrami-Laplace operator, a generalization of the Laplacian for the $k$-forms, which is expressed in
 local coordinates by the Weitzenbock formula \cite{Laplace}
\be
(\Delta_k\omega)_{i_1..i_k}=-\nabla^2 \omega_{i_1..i_k}+\sum_{\nu=1}^k (-1)^\nu R^n_{\ i_\nu}\omega_{ni_1..\hat{i}_\nu..i_k}+2\sum_{\mu<\nu}(-1)^{\mu+\nu}R^{n\ m}_{\ i_\nu\ i_\mu}\omega_{mni_1..\hat{i}_\mu..\hat{i}_\nu..i_k}\, .
\lb{Deltap}
\ee
It takes the form $-\nabla^2+X^{(k)}$, where the exact form of $X^{(k)}$ can be deduced from (\ref{Deltap}). The calculation shows that $\tr X^{(1)}=R$, $\tr X^{(2)}=(d-2)R$,
$\tr X^{(3)}=\frac{(d-2)(d-3)}{2}R$. The natural extrapolation to arbitrary $k$ is $\tr X^{(k)}=\frac{(d-2)..(d-k)}{(k-1)!}R=C^{k-1}_{d-2}\, R$. 
In dimension $d=k$ the $k$-form is dual to 
$0$-form and thus the operator $\Delta_k$ becomes the usual Laplace operator acting on scalar functions. Therefore $\Tr X^{(k)}=0$ if $d=k$ for $k>1$.

The heat kernel coefficient  corresponding to the product of determinants in (\ref{Zp}) is a sum of coefficients for operators (\ref{Deltap}) of various rank,
\be
a^{(p)}_1=\sum_{k=0}^p(k+1)(-1)^k\, a_1(\Delta_{p-k})\, .
\lb{ak}
\ee
For the heat kernel coefficient of  operator $\Delta_k$ we have the formula (\ref{a2-general spin})
\be
a_1(\Delta_{k})=\int d^dx\sqrt{g}\left(\frac{D_k(d)}{6}-\Tr X^{(k)}\right)\, , \, \, D_k(d)=C^k_d\, , \, \, \Tr X^{(k)}=C^{k-1}_{d-2}R\, ,
\lb{A1k}
\ee
where we take into account that the number of independent components of an antisymmetric tensor of rank $k$ in $d$ spacetime dimensions is $C^k_d$.

Now using the identity
\be
\sum_{k=0}^pC^{p-k}_d(-1)^{k}(k+1)=C^p_{d-2}\, 
\lb{indentity}
\ee
the sum in (\ref{ak}) can be evaluated and we find that\footnote{As we have learned this result was earlier obtained in \cite{Fradkin:1982kf}.}
\be
a^{(p)}_1=\int d^d x\sqrt{g}\left(\frac{{\cal D}_p(d)}{6}-c^{(p)}(d)\right)R \, ,
\lb{app}
\ee
where ${\cal D}_p(d)=C^p_{d-2}$ is number of on-shell components of $p$-form and
\be
c^{(p)}(d)=C^{p-1}_{d-4}\, 
\lb{cp}
\ee
is the corresponding contact term. This expression generalizes Kabat's contact term (\ref{cs1})  to arbitrary gauge $p$-form.
Notice that ${\cal D}_p(d)$ as function of dimension $d$ is a polynome of degree $p$ while $c^{(p)}(d)$ is polynome of degree $(p-1)$.

Using the relation $p(d-2-p)C^p_{d-2}=(d-2)(d-3)C^{p-1}_{d-4}$ we can re-write (\ref{app}) as
\be
a^{(p)}_1=C^{p-1}_{d-4}\int d^d x\sqrt{g}(\frac{(d-2)(d-3)}{6p(d-2-p)}-1)R\, .
\lb{aR}
\ee
This shows that the corresponding contribution to the induced Newton constant is mostly negative for $p\leq d-3$. The only case (for $p>1$) when 
it exactly vanishes is $d=8$, $p=5$.

\section{Graviton, $s=2$}
\setcounter{equation}0

Our next example is the graviton. This, obviously very important case,  is less studied in the literature. Therefore, we shall give as much details as possible.
The first important question here is what the appropriate differential operator which describes the quantum dynamics of the graviton?
In the literature one typically uses the Lichnerowicz operator. However, the Lichnerowicz operator is not the one who governs the dynamics of the quadratic 
fluctuations in the Einstein action. In order to find the form of the appropriate operator we shall study the quadratic term in the Einstein action 
\be
W_{EH}=-\frac{1}{16\pi G}\int d^dx\sqrt{g}R
\lb{EH}
\ee
with respect to 
perturbation of the metric, $g_{\mu\nu}=g^{(0)}_{\mu\nu}+\sqrt{16\pi G}h_{\mu\nu}$, where $g^{(0)}_{\mu\nu}$ is the background metric and $h_{\mu\nu}$ is the perturbation.
We stress that we want to see what  Newton constant is induced by these fluctuations. Therefore, the background metric is not supposed to be constrained 
by any field equations. In what follows we shall skip the script $(0)$ for the background metric.

\subsection{Linear  perturbations and the Lichnerowicz operator}
To start with let us consider the linear perturbation of the Ricci tensor,
\be
&&\delta_1 R_{\mu\nu}=\frac{1}{2}(\nabla^\alpha\nabla_\mu h_{\nu\alpha}+\nabla^\alpha\nabla_\nu h_{\mu\alpha}-\nabla^2 h_{\mu\nu}-\nabla_\mu\nabla_\nu h)\nonumber \\
&&=\frac{1}{2}\nabla_\mu \nabla_\alpha(h^\alpha_\nu-\frac{1}{2}\delta^\alpha_\nu h)+\frac{1}{2}\nabla_\nu \nabla_\alpha(h^\alpha_\mu-\frac{1}{2}\delta^\alpha_\mu h)+\frac{1}{2}\Delta_L h_{\mu\nu}\, ,
\lb{Ricci}
\ee
where in the second line we commuted the covariant derivatives and $\Delta_L$ is the Lichnerowicz operator
\be
\Delta_L h_{\alpha\beta}=-\nabla^2 h_{\alpha\beta}+R_{\alpha}^{\ \mu}h_{\mu\beta}+R_{\beta}^{\ \mu}h_{\mu\alpha}-2R^{\mu\ \nu}_{\ \alpha\ \beta}h_{\mu\nu}\, .
\lb{L}
\ee 
Assume that the linear perturbation   
 satisfies the gauge condition
 \be
 \nabla^\alpha(h_{\alpha\beta}-\frac{1}{2}g_{\alpha\beta}h)=0\, , \,\, h=g^{\mu\nu}h_{\mu\nu}\, .
 \lb{h}
 \ee
 one finds that the  linear perturbation of the Ricci tensor 
 \be
 \delta_1 R_{\mu\nu}=\frac{1}{2}\Delta_L h_{\mu\nu}\, .
 \lb{R1}
 \ee
 is governed by the Lichnerowicz operator.
In fact, the Ricci tensor is the only tensor the small deformation of which is determined by the Lichnerowicz operator without any curvature dependent modifications. 

  Respectively, the linear perturbation of the Einstein tensor $G_{\mu\nu}=R_{\mu\nu}-\frac{1}{2}Rg_{\mu\nu}$
  reads
  \be
 && \delta_1 G_{\mu\nu}=\frac{1}{2}\nabla_\mu\nabla_\alpha(h^\alpha_\nu-\frac{1}{2}\delta^\alpha_\nu h)+\frac{1}{2}\nabla_\nu\nabla_\alpha(h^\alpha_\mu-\frac{1}{2}\delta^\alpha_\mu h)
 -\frac{1}{2}\nabla^\alpha\nabla^\beta (h_{\alpha\beta}-\frac{1}{2}g_{\alpha\beta} h)\nonumber \\ 
 && +\frac{1}{2}\Delta_L h_{\mu\nu}+\frac{1}{4}g_{\mu\nu}\nabla^2 h-\frac{1}{2}h_{\mu\nu}R+\frac{1}{2}g_{\mu\nu}h_{\sigma\rho}R^{\sigma\rho}
  \lb{G}
  \ee
 In the gauge (\ref{h}) the first line in (\ref{G}) vanishes and we have
 \be
  \delta_1 G_{\mu\nu}=\frac{1}{2}\Delta_L h_{\mu\nu}+\frac{1}{4}g_{\mu\nu}\nabla^2 h-\frac{1}{2}h_{\mu\nu}R+\frac{1}{2}g_{\mu\nu}h_{\sigma\rho}R^{\sigma\rho}\, .
  \lb{G-1}
  \ee
  It is useful to introduce a traceless part of the perturbation, $\phi_{\mu\nu}=h_{\mu\nu}-\frac{1}{d}g_{\mu\nu}h$.  Decomposing the Einstein tensor on the trace and traceless parts,
  \be
  \delta_1 G=g^{\alpha\beta}\delta_1 G_{\alpha\beta}\, , \,\, \,\,\delta_1 G^{TT}_{\alpha\beta}=\delta_1 G_{\alpha\beta}-\frac{1}{d}g_{\alpha\beta}  \delta_1 G
  \lb{GG}
  \ee
  we find 
  \be
 && \delta_1 G=\frac{(d+2)}{4}\nabla^2 h +\frac{d}{2}\phi_{\mu\nu}R^{\mu\nu}\, ,\nonumber \\
 &&\delta_1 G^{TT}_{\alpha\beta}=\frac{1}{2}\Delta_L\phi_{\alpha\beta}-\frac{1}{2}\phi_{\alpha\beta}R\, .
 \lb{GG1}
 \ee
If the background spacetime is Einstein, i.e. $R_{\mu\nu}=\Lambda g_{\mu\nu}$, then $\phi_{\mu\nu}R^{\mu\nu}=0$ and the two modes (the trace and the traceless one)
decouple.  But even in this case the relevant operator, which governs the linear transverse traceless perturbations of the Einstein metric, in the Einstein equations is operator $(\Delta _L-R)$ which is a deformation of  the Lichnerowicz operator by a curvature dependent term. We shall see in the next section that the deviation from $\Delta_L$ is even more important if
we analyze the operator for the quadratic perturbations in the Einstein-Hilbert action. This consideration shows that the Lichnerowicz operator without any curvature dependent modifications appears only in the perturbation of the Ricci tensor. For other tensors, such as the Einstein tensor, the modifications due to some curvature dependent terms is necessary.
Since the curvature dependent terms in  $X^{(s=2)}$ are essential for the structure of the induced Newton constant we conclude that it is not reasonable to consider the operator 
$\Delta_L$ alone as the appropriate operator for the spin $s=2$ particles. The curvature dependent modifications of $\Delta_L$ should be considered that necessarily leads to 
modification in the induced Newton constant caused by the spin $s=2$ particles.

\subsection{Quadratic perturbations in Einstein-Hilbert action}
Let us now analyze what operator governs the quadratic perturbations in  the Einstein-Hilbert action. For this we first consider the term quadratic in perturbations 
in the gravitational action (\ref{EH}),
\be
&&\delta_2W_{EH}=-\int d^dx\sqrt{g}\nabla_\mu(h^{\mu\nu}-\frac{1}{2}g^{\mu\nu}h)\nabla_\alpha(h^\alpha_\nu-\frac{1}{2}\delta^\alpha_\nu h)\nonumber \\
&&-\int d^dx\sqrt{g}[\frac{1}{2}h^{\alpha\beta}\Delta_L h_{\alpha\beta}+\frac{1}{4}h\nabla^2 h-2h_{\mu}^{\ \nu}G_{\nu}^{\ \alpha}h_{\alpha}^{\ \mu}-\frac{1}{4}h^2 R+h h_{\alpha\beta}R^{\alpha\beta}-\frac{1}{2}h_{\alpha\beta}h^{\alpha\beta}R]\, ,
\lb{W2}
\ee
where $G_{\alpha\beta}=R_{\alpha\beta}-\frac{1}{2}g_{\alpha\beta}R$ is the Einstein tensor and we raise  indexes with the help of the background metric $g^{\alpha\beta}$.
The term in the first line in (\ref{W2}) originates from the first line in (\ref{G}).
As 't Hooft and Veltman \cite{'tHooft:1974bx} have instructed us long ago it is convenient to add  the following  gauge fixing term
\be
W=\delta_2W_{EH}+W_{gauge}\, , \, \, W_{gauge}=\int d^d x\sqrt{g}\nabla_\mu(h^{\mu\nu}-\frac{1}{2}g^{\mu\nu}h)\nabla_\alpha(h^\alpha_\nu-\frac{1}{2}\delta^\alpha_\nu h)
\lb{fix}
\ee
that cancels the term in the first line in (\ref{W2}).
As in the previous section, it is useful to introduce the traceless perturbation $\phi_{\alpha\beta}=h_{\alpha\beta}-\frac{1}{d}g_{\alpha\beta}h$. Then the quadratic action
(\ref{fix}) is expressed in terms of $\phi_{\alpha\beta}$ and $h$ 
\be
&& W'=-\int d^dx\sqrt{g}[\frac{1}{2}\phi_{\alpha\beta}\Delta_L\phi^{\alpha\beta}-2\phi_\mu^{\ \nu}(R_{\nu}^{\ \alpha}-\frac{1}{4}\delta_\nu^\alpha R)\phi_\alpha^{\ \mu}\nonumber \\
&&+\frac{(d-4)}{d}h\phi_{\alpha\beta}R^{\alpha\beta}+\frac{(2-d)}{4d}(-h\nabla^2 h+\frac{(d-4)}{d}Rh^2)]\, .
\lb{W22}
\ee
In dimension $d=4$ one reproduces the earlier results  \cite{Gibbons:1978ji}, \cite{Christensen:1979iy}.

Some remarks are in order.

\noindent 1. First of all it is important to note that the conformal mode $h$ becomes non-minimally coupled if $d\neq 4$. 
The corresponding operator takes the form (\ref{diff-operator})  
\be
(-\nabla^2 +X^{(h)})\, , \, \, X^{(h)}=\frac{(d-4)}{d}R\, .
\lb{non-min}
\ee
This non-minimal coupling disappears if $d=4$ but otherwise is generically present. To the best of our knowledge this fact is not known in the literature.

\noindent 2. There generically presents an off-diagonal coupling between $\phi_{\alpha\beta}$ and $h$. This coupling disappears only if $d=4$.

\noindent 3. In the traceless transverse sector we see the appearance of an extra term $\phi_\mu^{\ \nu}(R_{\nu}^{\ \alpha}-\frac{1}{4}\delta_\nu^\alpha R)\phi_\alpha^{\ \mu}$.
This extra term disappears in $d=4$ if the background metric is Einstein so that the differential operator acting on the  transverse traceless mode $\phi_{\alpha\beta}$ is 
the Lichnerowicz operator. In dimensions $d>4$ this terms does not disappear even if the metric is Einstein (provided the scalar curvature is non-vanishing).
Thus,  in higher dimensions   an important curvature dependent modification of the Lichnerowicz operator appears.

The above observations show that $d=4$ case is in many respects special. However, even in $d=4$ for a generic (non-Einstein) background metric the operator in the
transverse traceless sector is not the genuine Lichnerowicz operator.

\subsection{Newton constant due to graviton in $d$ dimensions }
The quadratic action (\ref{fix}) should be supplemented with the ghost term which corresponds to our choice of the gauge fixing,
\be
W=W'+W_{ghost}\, , \, \, W_{ghost}=\frac{1}{2}\int d^d x\sqrt{g}~\chi_\mu(-g^{\mu\nu}\nabla^2 -R^{\mu\nu})\chi_\nu\, .
\lb{ghost}
\ee
The corresponding ghost contribution  to $a_1$ is
\be
a_1^{gh}=\frac{d}{6}R-\tr X_{gh}\, , \,\, \tr X_{gh}=-R\, .
\lb{a-gh}
\ee

\medskip

As it is seen from the structure of the action (\ref{W22}) the trace and traceless sectors should be considered together as a composite field
$(\phi_{\mu\nu}, \, h)$. The corresponding kinetic terms\footnote{The kinetic term for the trace mode has a wrong sign. We use the standard prescription $h\rightarrow i h$.} however diagonalize in each sub-sector of $\phi_{\mu\nu}$ and $h$. So that only the $X$-matrix
has the off-diagonal $(\phi_{\mu\nu} h)$ components. These components however are not important when we compute the trace of the complete matrix $X$ in the $(\phi, h)$ sector,
\be
\Tr X^{(\phi,h)}=\tr X^{TT}+\tr X^{(h)}\, ,
\lb{XXh}
\ee
where $\tr X^{(h)}$ is found in (\ref{non-min}) and matrix $ X^{TT}$ is defined in the traceless sector.
In this sector  we have
\be
\frac{1}{2}\int d^d x\sqrt{g}~\phi_{\alpha\beta}\left(-\delta_{\mu\nu}^{\alpha\beta}\nabla^2 +X^{TT, \alpha\beta}_{\mu\nu}\right)\phi^{\mu\nu}\, ,
\lb{TTT}
\ee
where 
\be
\delta_{\mu\nu}^{\alpha\beta}=\frac{1}{2}(\delta^\alpha_\mu\delta^\beta_\nu+\delta^\beta_\nu\delta^\alpha_\mu)-\frac{1}{d}g_{\mu\nu}g^{\alpha\beta}
\lb{projector}
\ee
is the projector on the traceless part of a symmetric tensor, its trace $\delta^{\alpha\beta}_{\alpha\beta}=\frac{d^2+d-2}{2}=D_{s=2}(d)$ is the number of components of a symmetric traceless tensor of rank 2.
Tensor  $X^{TT}_{\alpha\beta,\mu\nu}$ is defined as 
\be
X^{TT}_{\alpha\beta,\mu\nu}=X_{\alpha\beta,\mu\nu}-\frac{1}{d}g_{\alpha\beta}X_{\sigma\rho,\mu\nu}g^{\sigma\rho}-\frac{1}{d}g_{\mu\nu}X_{\alpha\beta,\sigma\rho}g^{\sigma\rho}
+\frac{1}{d^2}g_{\mu\nu}g_{\alpha\beta}X_{\sigma\rho,\epsilon\kappa}g^{\sigma\rho}g^{\epsilon\kappa}\, ,
\lb{XTT}
\ee
where $X_{\alpha\beta,\mu\nu}$ takes the form
\be
X_{\alpha\beta,\mu\nu}=-R_{\mu\alpha\nu\beta}-R_{\nu\alpha\mu\beta}-\frac{1}{2}(R_{\alpha\nu}g_{\beta\mu}+R_{\alpha\mu}g_{\beta\nu}+R_{\beta\nu}g_{\alpha\mu}+R_{\beta\mu}g_{\alpha\nu})+\frac{R}{2}(g_{\mu\alpha}g_{\nu\beta}+g_{\mu\beta}g_{\nu\alpha})
\lb{X}
\ee
The trace of (\ref{XTT}) is evaluated as
\be
X^{TT,\alpha\beta}_{\alpha\beta}=X^{\alpha\beta}_{\ \  \alpha\beta}-\frac{1}{d}X_{\mu\nu,\alpha\beta}g^{\mu\nu}g^{\alpha\beta}\, .
\lb{TrX}
\ee
We find that 
\be
X^{\alpha\beta}_{\ \  \alpha\beta}=R~\frac{d(d-1)}{2}\, , \, \, X_{\mu\nu,\alpha\beta}g^{\mu\nu}g^{\alpha\beta}=R\,(d-4)
\lb{XXX}
\ee
and hence
\be
\tr X^{TT}=X^{TT,\alpha\beta}_{\ \  \ \ \alpha\beta}=R~(\frac{d^2-d-2}{2}+\frac{4}{d})\, .
\lb{XTTT}
\ee
The coefficient $a_1$ in the sector $(\phi,h)$ then
\be
a_1^{(\phi, h)}=\frac{D_{s=2}(d)+1}{6}R-\Tr X^{(\phi,h)}\, , \,\, \Tr X^{(\phi,h)}=\frac{d(d-1)}{2}\, .
\lb{aTh}
\ee

The above analysis in the traceless sector  can be compared with the analysis for the Lichnerowicz operator (\ref{L}). The matrix $X$ in this case is
\be
X^{L,\mu\nu}_{\ \ \ \alpha\beta}=-R^{\mu\ \nu}_{\ \alpha \ \beta}-R^{\nu\ \mu}_{\ \alpha\ \beta}+\frac{1}{2}(R_{\alpha}^{\ \mu}\delta^\nu_\beta+(\mu\leftrightarrow\nu, \alpha\leftrightarrow\beta))\, .
\lb{XL}
\ee
Note, that it is already traceless, $X^{L,\mu\nu}_{\ \ \ \alpha\beta}g^{\alpha\beta}=0$ and $X^{L,\mu\nu}_{\ \ \ \alpha\beta}g_{\mu\nu}=0$. Then we find that
\be
\tr X^L=(d+2)R\, .
\lb{XLL}
\ee
Curiously, (\ref{XLL}) and (\ref{XTTT}), being generically quite different functions of dimension $d$, coincide if $d=4$. This is due to the fact the difference of the matrices (\ref{XTT}) and (\ref{XL}) although non-vanishing has  zero trace  if $d=4$. 

Returning to our analysis of the renormalization of the Newton constant due to the graviton\footnote{The earlier works, where a  relevant analysis in dimensions $d=4$ and higher was present, include
\cite{Fradkin:1978yf}, \cite{Fradkin:1981iu}, \cite{Fradkin:1982kf}. In particular,  in section 4.1 of \cite{Fradkin:1982kf} it is given a calculation similar to the one presented here.
However, instead of separating the gravitational perturbations on the traceless and the trace parts as we do the authors of  \cite{Fradkin:1982kf} deal with the operator acting on the complete gravitational tensor modes. The relevant operator is the one which appears in  (\ref{W2}), (\ref{fix}) and it has the  form $(-P\nabla^2+X)$ where $P$ and $X$ are given
in (4.5) of \cite{Fradkin:1982kf}. This operator factorizes on a product of $P$ and $(-\nabla^2 \, I+P^{-1}X)$ so that only the latter produces an essential contribution to the
determinant. The heat kernel coefficient is $a_1=\Tr(R/6 I-P^{-1}X)$. One then finds that $\Tr I=d(d+1)/2$ and $\Tr P^{-1}X=R d(d-1)/2$ that exactly  corresponds to our equation
(\ref{aTh}). So that this way we again reproduce our results (\ref{aTh}) and, taking into account the ghost contribution,  (\ref{as2}).  The result given in (4.6) of \cite{Fradkin:1982kf} is, however, different from (\ref{aTh}) by $(d-4)/2$ and, as  we believe,  is incorrect. We thank A. Tseytlin for bringing to us his earlier relevant work
and for the discussion of the relation between our result and the one given in \cite{Fradkin:1982kf}.}, we find that the complete coefficient $a^{(s=2)}_1$ is the sum of contributions of
the $(\phi,h)$ sector and of the ghost  (\ref{a-gh}),
\be
&&a_1^{(s=2)}=a_1^{(\phi,h)}-2a_1^{gh}=\frac{1}{6}(D_{s=2}+1-2d)R-(\frac{d^2-d-2}{2}+\frac{4}{d}+\frac{(d-4)}{d}+2)R\nonumber \\
&&=\frac{{\cal D}_{s=2}(d)}{6}R-\frac{(d^2-d+4)}{2}R\, ,
\lb{as2}
\ee
where ${\cal D}_{s=2}(d)=\frac{d(d-3)}{2}$ is number of on-shell degrees of freedom of graviton in dimension $d$.
Notice, that in this expression the non-polynomial term $4/d$ of (\ref{XTTT}) cancels with $-4/d$ coming from (\ref{non-min}). 
The contribution to the induced Newton constant thus takes the form (\ref{Gren-spin-s}) with the contact term 
\be
c_{(s=2)}(d)=\frac{d^2-d+4}{2}\, .
\lb{cs2}
\ee
This result (without details) is given in \cite{Solodukhin:2011gn}.
Notice that the complete contribution to the Newton constant due to graviton is  negative in any dimension $d\geq 4$.

\section{Some comments }
\setcounter{equation}0

Let us make some comments.

\medskip

\noindent{\bf 1.} Our first observation is that the induced Newton constant is not positive in general. Indeed, this is clear from eq. (\ref{Gren-spin-s}).
In fact this was already noticed by Adler in his review \cite{Adler:1982ri}.  This may look surprising that the theory does not automatically  predict the correct sign for the
the strength of the most important interaction. Curiously enough this sign is not fixed by the general consistency requirements such as unitarity.
Thus, the positivity, if required,  may be only achieved by an appropriate choice of the fields multiplet.

\medskip

\noindent{\bf 2.} Even though $1/G(\epsilon)$ is not positive it can be presented as a difference of two apparently positive quantities. The first one
is the first term in (\ref{Gren-spin-s}). It represents the contribution of the physical degrees of freedom to the renormalization of the Newton cosntant.
This contribution is obviously positive: each physical degree of freedom, no matter bosonic or fermionic, 
contributes positively   to $1/G(\epsilon)$. The second quantity is what we call the ``contact term'' $c^{(s)}$.  It comes from the trace of $X^{(s)}$ in the field operator.
We do not have a general proof that $c^{(s)}$ should be positive. However, the simple inspection of the examples considered  shows that
in  these cases $c^{(s)}$ is positive. The only possible exception is the scalar field for which the sign of the contact term depends on the sign of the non-minimal coupling $\xi$. In fact, it appears that the negative values of $\xi$ for the Higgs boson are phenomenologically preferred \cite{Higgs}.

\medskip

\noindent{\bf 3.} In certain cases $1/G(\epsilon)$ may vanish. This is so in the case of the conformal scalar in $d=4$. However, the vanishing of $1/G(\epsilon)$ is not a general consequence of the conformal
symmetry. The only other conformal field theory  we know with same property in four dimensions  is the ${\cal N}=4$ $SU(N)$  super-Yang-Mills theory.
The respective matter multiplet contains 6 conformal scalars, 2 Dirac fermions and 1 vector field.  

\medskip

\noindent{\bf 4.} The observable division of all fields in the Universe on the matter fields and those  that mediate their interactions (gauge bosons, gravitons)
is particularly sharp when we discuss their respective contributions to the Newton constant. Only the fields-mediators have the non-vanishing contact terms and thus produce
the negative contributions to the Newton constant. The Higgs boson might have some non-trivial  non-minimal coupling to gravity and in this respect it might be similar to the mediators.

\medskip

\noindent{\bf 5.}  One might worry that the contact term (\ref{cs2}) due to graviton is quadratic function  of dimension  $d$ compared to the one due to
a gauge vector field which does not depend on $d$. However, the dependence of the contact terms on $d$ should be compared with such a dependence of the number
of the off-shell degrees of freedom. If we do so we find that both $s=2$ and $s=1$ cases follow the same pattern:
\be
D_s(d)-c^{(s)}(d)=d-s\, , \, \, s=1, \, 2\, ,
\lb{Dc}
\ee
where $D_{s=1}(d)=d$ and $D_{s=2}(d)=\frac{d(d+1)}{2}$ is the number of components of symmetric tensor of rank 2.

\medskip

\noindent{\bf 6.} In the Einstein-Maxwell theory the vector and gravitational perturbations do not decouple (this case was considered by Sen in \cite{Sen}, where the main focus was made to the logarithmic terms). The relevant second order operator is not diagonal
and the result for the effective action  does not simply factorize on the sum of the contributions due to spins $s=1$ and $s=2$. 
Nevertheless we expect that the factorization takes place for the Newton constant and the area law in the entropy. The details, however, would be interesting to work out.

\medskip

\noindent{\bf 7.} An important question is whether the contact terms are gauge independent. The analysis given in 5th ref. in \cite{contact}
shows that in the case of gauge vector fields ($s=1$) the contact term does not depend on the choice of the gauge fixing term.
However, for gravitons, $s=2$, it is known that the $R$-term in the effective action may depend on the gauge, see for instance \cite{Kallosh:1978wt}, \cite{Fradkin:1983nw}
(see also \cite{Falls:2015qga}).
Since the part due to the number of the on-shell degrees of freedom is gauge independent  this means that the contact term due to spin $s=2$ may
depend on the gauge.

\section{Entropy}
\setcounter{equation}0

Let us now discuss the entropy. It is defined for a co-dimension two surface $\Sigma$ by applying the so-called replica method.
The main case we are interested in is when the surface $\Sigma$ is a black hole horizon. Therefore, for simplicity we shall assume that 
in the curved spacetime there exists a timelike Killing vector $\partial_\tau$ such that it is null at the horizon surface $\Sigma$.
The various generalizations are of course possible: when the surface $\Sigma$ is a generic surface (not a fixed point of any isometry)
characterized by some extrinsic curvature, when the Killing symmetry is only approximative, valid in a small vicinity of $\Sigma$ and many others.
In the replica method one identifies the time coordinate $\tau$ with period $2\pi \alpha$ and computes the path integral for a field in question
on a manifold with this new periodicity. The respective quantum effective action is function of $\alpha$, $W(\alpha)$.
We are in particularly interested in the case when $\alpha=n$ is an integer. The partition function then  can be identified with a trace of  $n$-th power of
the reduced density matrix,
\be
e^{-W(\alpha=n)}=\Tr \rho^n\, .
\lb{rho}
\ee
The entropy then is defined by analytically extending (\ref{rho}) to non-integer values of $\alpha$ and computing the derivative with respect to $\alpha$,
\be
S(\alpha)=(\alpha\partial_\alpha-1)W(\alpha)|_{\alpha=1}\, .
\lb{S}
\ee
For integer $\alpha=n>1$ the Euclidean manifold in question is an $n$-fold cover of the original spacetime defined at $n=1$. It has a conical singularity with angle deficit
$\delta=2\pi(1-\alpha)$ at the surface $\Sigma$. Only this singularity contributes to the entropy (\ref{S}). 

Technically it is more convenient to work with 
the case when $\alpha=1/N<1$. The manifold in this case is an orbifold.  Namely this space arises when we consider the gravitational field created by a cosmic string of positive tension. Geometrically and topologically these two manifolds, the orbifold and $n$-fold cover, are quite different. 
The conical singularities however in these two cases  are analytically related and to some extent one may not worry about the  differences between them.
The attention however is required when the higher spins are considered. Then some non-analyticity in passing from $\alpha<1$ to $\alpha>1$
may arise. We shall discuss this issue later in the paper.

As we have seen the partition function for a field of spin $s$ reduces to calculation of determinant of some Laplace type operators
\be
-\nabla^2\delta_{AB}+X^{(s)}_{AB}\, ,
\lb{Operator}
\ee
where the potential term $X^{(s)}_{AB}$ is constructed from the Rieamann tensor,
\be
X^{(s)}_{AB}=R_{\alpha\beta\mu\nu}U_{AB}^{\alpha\beta\mu\nu}\, .
\lb{X}
\ee
Thus, the calculation of the effective action (\ref{rho}) and the entropy (\ref{S}) reduces to calculation of the heat kernel for the operator
of the type (\ref{Operator}) on a manifold with a conical singularity. The main question here is how to treat the potential term (\ref{X}).
On a conical manifold the Riemann curvature has a delta-like contribution due to the singularity \cite{Fursaev:1995ef}
\be
R_{\alpha\beta\mu\nu}=2\pi(1-\alpha)((n_\alpha, n_\mu)(n_{\beta},n_\nu)-(n_\alpha,n_\nu)(n_\beta n_\mu))\delta_\Sigma+R_{\alpha\beta\mu\nu}^{reg}\, ,
\lb{Riemann}
\ee
where $(n_\alpha, n_\beta)=n_\alpha^a n_\beta^a$, $n^a_\mu, \, a=1,2$ is a pair of vectors normal to $\Sigma$,  $R_{\alpha\beta\mu\nu}^{reg}$ is the regular part of the curvature
which remains when $\alpha=1$.

The question now is whether we should  take into account this singular term in the curvature (\ref{Riemann}) when compute the effective action and the entropy or this term should be ignored? There are two possible  answers to this question and, respectively,  two different  entropies can be defined.

\subsection{Entanglement entropy}

One possibility is to treat the potential term (\ref{X}) as any other potential term. This suggests the following prescription.   First we consider 
operator (\ref{Operator}) with some unspecified potential
\be
\hat{\textbf{D}}=-\nabla^2\delta_{AB}+V_{AB}\, ,
\lb{Operator1}
\ee
compute the partition function $Z(\alpha)$ for this operator on a conical space, deduce the entropy $S=-(\alpha\partial_\alpha-1)\ln Z(\alpha)|_{\alpha=1}$
and only at the end make the substitution: $V_{AB}=X^{(s)}_{AB}$. In this way we completely ignore the singular term in (\ref{X}) due to the singularity in the Riemann tensor 
(\ref{Riemann}). This prescription is motivated by that simple  observation that the construction of the reduced density matrix from a vacuum state
in Minkowski space-time does not appear to be  affected by the presence of the Riemann tensor in the potential term (\ref{X}). Indeed in flat spacetime this potential terms
is vanishing and the reduced density matrix appears to be exactly the same as for an operator with $X^{(s)}=0$. 
The entropy defined by this prescription we shall call the ``entanglement entropy''. It corresponds to a well-defined density matrix and, in particular,
this entropy is positive definite as we show below. 

 The calculation of the partition function $Z(\alpha)$ and the entropy $S(\alpha)$ can be performed using the heat kernel technique.  The coefficients in the expansion 
 (\ref{K expansion}), being considered on a conical space, are modified by certain surface terms,
 \be
 a_n(\alpha)=\alpha a_n(\alpha=1)+a_n^{\Sigma}(\alpha)\, ,
 \lb{ans}
 \ee
 where $a_n(\alpha=1)$ are the coefficients in the heat kernel expansion on regular spacetime and $a_n^{\Sigma}(\alpha)$ are the surface terms given by integrals over the 
 entangling surface $\Sigma$. In this paper, for the analysis of the leading divergences of the entropy,  we shall need only the first $n=1$ coefficient. For an operator of the type (\ref{Operator1}) we have that
 \be
 a_1^{\Sigma}(\alpha)=\frac{\pi}{3}\frac{(1-\alpha^2)}{\alpha}(-1)^F D(d)\int_\Sigma 1\, ,
 \lb{a1s}
 \ee
 where $D(d)=\delta_{AA}$ is the dimension of the representation. If the operator (\ref{Operator1}) is acting on fields of half-integer spin then the surface term (\ref{a1s}) is negative compared to the case of fields of integer spin, this behavior is taken care by the factor $(-1)^F$ in (\ref{a1s}).
 Notice that the first coefficient (\ref{a1s}) does not depend on the potential $V$.
The case of the Dirac fermions was first considered in \cite{Larsen:1994yt}. Here we closely follow the discussion in  \cite{Fursaev:1996uz}.

It should be noted that for spins $s=2$ and $s=3/2$ there have been found  in \cite{Fursaev:1996uz} (see also \cite{He:2014gva}) some extra surface terms additional to (\ref{a1s}) which do not disappear
when $\alpha=1$. These non-analytic terms are not written in (\ref{a1s}). They will be discussed later in section \ref{orbi} where we argue why these terms should not be present in the $n$-fold cover construction, i.e. when $\alpha=n>1$.

In the examples considered above the complete partition function for a field of spin $s$ contains a product of positive and negative powers of determinants
of the Laplace type operators $\hat{\textbf{D}}_k$ (\ref{Operator1}),
\be
Z^{(s)}=\prod_k {\det}^{-\nu(k)/2}\hat{\textbf{D}}_k\, ,
\lb{Dk}
\ee
where $\nu(k)$ are some numbers, positive or negative. For each operator $\hat{\textbf{D}}_k$ one has the heat kernel expansion (\ref{ans}), (\ref{a1s}).
The leading UV divergence of the surface part of the effective action then can be computed as
\be
W^\Sigma(\alpha)=\frac{1}{d-2}\frac{1}{(4\pi)^{d/2}}\frac{1}{\epsilon^{d-2}}\frac{\pi}{3}\frac{(1-\alpha^2)}{\alpha}\, \sum_k\nu_k\, (-1)^{F}D_k(d)\, \int_\Sigma 1\, ,
\lb{Wk}
\ee
where $D_k(d)$ is the dimension of the representation in the operator $\hat{\textbf{D}}_k$. If the operator $\hat{\textbf{D}}_k$ is acting on fields of half-integer spin its contribution in (\ref{Wk}) is counted with negative sign.   In all cases considered in this paper we have observed that the sum
\be
\sum_k\nu_k(-1)^{F}D_k(d)={\cal D}_s(d)\,
\lb{Dkk}
\ee
is the number of on-shell degrees of freedom of spin $s$.  Using this property and the form (\ref{Wk}) for the surface action we 
find the leading term in the entanglement entropy due to spin $s$, 
\be
S_{ent}^{(s)}={1\o (4\pi)^{d-2 \o 2}(d-2)}  {1\o \epsilon^{d-2}}\frac{{\cal D}_{s}(d)}{ 6}A(\Sigma)\, ,
\lb{Sent}
\ee
$A(\Sigma)$ is the area of surface $\Sigma$. We see that only physical degrees of freedom contribute to this entropy. Moreover, the entropy (\ref{Sent}) is obviously positive
as it should be provided the entropy is determined by a well-defined density matrix. The other observation is that the entanglement entropy  (\ref{Sent}) is gauge independent.

The generalization to the massive fields is straightforward and is carried by the multiplication of the heat kernel by a factor $e^{-m^2\tau}$.  The complete result for the entanglement entropy due to a set of spin $s$ and mass $m_s$ is
\be
S_{ent}=\frac{1}{2}\sum_s  \frac{{\cal D}_{s,m}(d)}{ 6} \int_{\epsilon^2}^\infty\frac{d\tau}{(4\pi\tau)^{d/2}}e^{-m_s^2\tau}\, A(\Sigma)\, .
\lb{Sents}
\ee
Comparing the area law (\ref{Sent}), (\ref{Sents}) and the renormalization of the Newton constant (\ref{Gren-spin-s}), (\ref{3.6})  we see that if there are no fields in the multiplet for which
the contact terms $c^{(s)}$ are non-vanishing then the entropy (\ref{Sent}) takes the Bekenstein-Hawking form
\be
S_{ent}=\frac{A(\Sigma)}{4G(\epsilon)}\, 
\lb{SS}
\ee
expressed in terms of the induced Newton constant $G(\epsilon)$. This is in fact a part of the theorem proved in \cite{Fursaev:1994ea}
about the simultaneous renormalization of entanglement entropy and the Newton constant.
In the presence of the gauge fields ($s=1$ and $s=2$) the situation, however, is different: due to the contact terms one can not in general write entanglement entropy
(\ref{Sent}) in the  form (\ref{SS}).

\subsection{Conical entropy}

The other possibility is to take into account the singularity in the curvature (\ref{Riemann}) even in flat spacetime. This reduces the problem to considering the quantum field theory
coupled to a delta-like potential. The complete treatment of the problem might be quite complicated (see for instance \cite{Solodukhin:1997xn}). We are however interested in the case when the angle deficit is very small so that the singular term in  (\ref{Riemann}) can be treated as a perturbation. For a non-minimal scalar this approach was suggested in 
\cite{Solodukhin:1995ak}. The perturbation theory for a delta-like potential is not well defined since the higher powers of the ``perturbation'' are singular non-integrable  functions.
This problem however is absent for the terms linear in the perturbation so that to linear order in $(1-\alpha)$ the perturbation theory  gives a meaningful result.
 The entropy which appears in this calculation  we shall
call the conical entropy since it is defined as a response to introduction of a small angle deficit in the effective action. The conical entropy is not related to a well-defined reduced density matrix and  it does not have, in general,  a clear statistical meaning. It has however other advantages which will be discussed below.

It should be noted that for the fermionic fields ($s=1/2,3/2$) the second order operator is obtained by squaring a first order Dirac type operator 
for which there is no a non-minimal coupling to gravity. Therefore, it appears reasonable that the singular terms in the curvature  should not be taken into account in the corresponding second order operators. Thus, below in this section we shall consider only bosonic operators.  

Considering the singular term in (\ref{Riemann}) as a perturbation to linear order we find a modification of the surface terms in the heat kernel coefficients of the operator (\ref{Operator}), (\ref{X}). This modification arises in exactly same way as in the case of the scalar field considered in \cite{Solodukhin:1995ak}. Generalizing this
consideration to a more general potential term (\ref{X}) and taking into account  (\ref{Riemann}) we find that to leading order in $(1-\alpha)$
\be
a_1^\Sigma=a_{1, (0)}^\Sigma-4\pi(1-\alpha)\int_\Sigma \Tr U^{\alpha\beta\mu\nu}(n_\alpha,n_\mu)(n_\beta,n_\nu)\, ,
\lb{con1}
\ee
 where $a_{1, (0)}^\Sigma$ is the unperturbed coefficient (\ref{a1s}), and
  the trace is taken with respect to indexes $AB$ of matrix $U$.  Now we notice that  the trace of the  potential term (\ref{X}) reduces to a Ricci scalar,
 \be
 \Tr X^{(s)}=\Tr U^{\alpha\beta\mu\nu}R_{\alpha\beta\mu\nu}=c\,R\, .
 \lb{con2}
 \ee
 For the Ricci scalar the formula (\ref{Riemann}) reduces to
 \be
 R=4\pi(1-\alpha)\delta_\Sigma+R^{reg}\, .
 \lb{con2-1}
 \ee
 Applying (\ref{Riemann}), (\ref{con2-1}) to both sides of (\ref{con2}) we find that
 \be
 \int_\Sigma \Tr U^{\alpha\beta\mu\nu}(n_\alpha,n_\mu)(n_\beta,n_\nu)=c \int_\Sigma 1\, 
 \lb{con3}
 \ee
 and the surface term in the heat kernel coefficient for a bosonic operator (\ref{Operator}) reads
 \be
 a_1^\Sigma=\left(\frac{\pi}{3}\frac{(1-\alpha^2)}{\alpha} D_s(d)-4\pi(1-\alpha)c\right)\int_\Sigma 1\, .
 \lb{con4}
 \ee
 
 \medskip
 
 Let us consider some examples.

\medskip

\noindent{\bf 1. Scalar operator} $-\nabla^2+\xi R$. This case was already considered in \cite{Solodukhin:1995ak}. In this case $c=\xi$ and we have for the surface coefficient
\be
 a_1^\Sigma=\left(\frac{\pi}{3}\frac{(1-\alpha^2)}{\alpha}-4\pi(1-\alpha)\xi\right)\int_\Sigma 1=4\pi(1-\alpha)(\frac{1}{6}-\xi)\int_\Sigma 1 +O(1-\alpha)^2\, .
 \lb{con5}
 \ee

\medskip

\noindent{\bf 2. Vector Laplace operator} $\Delta_1=-\nabla^2\delta_{\mu\nu}+R_{\mu\nu}$. In this case $c=1$ and we have that
  \be
 a_1^\Sigma=\left(\frac{\pi}{3}\frac{(1-\alpha^2)}{\alpha} d-4\pi(1-\alpha)\right)\int_\Sigma 1\, .
 \lb{con6}
 \ee

\medskip

\noindent{\bf 3. Laplace operator on $p$-forms}  $\Delta_p=(d\delta+\delta d)$  (\ref{Deltap}). In this case $c=C^{p-1}_{d-2}$ and hence
\be
a_1^\Sigma=\left(\frac{\pi}{3}\frac{(1-\alpha^2)}{\alpha} C^p_d-4\pi(1-\alpha)C^{p-1}_{d-2}\right)\int_\Sigma 1\, . 
 \lb{con7}
 \ee

\medskip

\noindent{\bf 4. Lichnerowicz operator} (\ref{L}) acting on symmetric tensors of rank 2: $c=(d+2)$ (as can be seen from (\ref{XLL})) and
\be
a_1^\Sigma=\left(\frac{\pi}{3}\frac{(1-\alpha^2)}{\alpha} \frac{d(d+1)}{2}-4\pi(1-\alpha)(d+2)\right)\int_\Sigma 1\, . 
 \lb{con8}
 \ee

\bigskip

These results for the vector and Lichnerowicz operators should be compared with the earlier results obtained by Fursaev and Miele \cite{Fursaev:1996uz} (and re-derived recently using some other methods in \cite{He:2014gva}). The analysis used by
these authors is based on exact spectrum of the operators and thus is supposed to be exact while our analysis above is based on a perturbative treatment of the singularity.
The comparison  shows that these results for the analytic terms (vanishing when $\alpha=1$) completely agree. This agreement for the Lichnerowicz operator however is not extended to the non-analytic term (independent of $\alpha$ and non-vanishing when $\alpha=1$)
found in  \cite{Fursaev:1996uz}.  We discuss the non-analytic terms in the next section.

We can now compute the surface terms in the effective action of the fields of spin $s$ considered earlier in the paper.
The replica formula (\ref{S}) applied to the surface effective action that corresponds to the surface heat kernel coefficients gives 
the following result for the leading area term due to a field of spin $s$ (where we have to take into account all contributions due non-minimal coupling in the  ghosts operators):
\be
S_{con}^{(s)}={1\o (4\pi)^{d-2 \o 2}(d-2)}  {1\o \epsilon^{d-2}}\left(\frac{{\cal D}_{s}(d)}{ 6}-c^{(s)}(d)\right)A(\Sigma)\, .
\lb{Scon}
\ee
Including the fields of various spin $s$ and mass $m_s$ we obtain 
\be
S_{con}=\frac{1}{2}\sum_s  \left(\frac{{\cal D}_{s,m}(d)}{ 6} -c^{(s,m)}(d)\right)\int_{\epsilon^2}^\infty\frac{d\tau}{(4\pi\tau)^{d/2}}e^{-m_s^2\tau}\, A(\Sigma)\, 
\lb{Scons}
\ee
for the complete conical entropy.
We see that this expression repeats exactly the renormalization of  the Newton constant (\ref{3.6}) so that we can write it in the form
\be
S_{con}=\frac{A(\Sigma)}{4G(\epsilon)}\, .
\lb{SG}
\ee
Namely this property of the conical entropy makes it so attractive. We, however, notice immediately that the conical entropy (\ref{Scon}) is not positive definite
and thus likely can not be obtained from a well-defined density matrix. The other observation is that the conical entropy (\ref{Scon}), (\ref{Scons}) is gauge  dependent 
since the contact term for graviton may depend on the gauge.

\subsection{Puzzle of non-analytic terms: orbifolds vs $n$-fold covers} 
\lb{orbi}

As we have already mentioned, Fursaev and Miele in their 1996 paper have found some non-analytic terms in the heat kernel of operator for spin $3/2$ and
of the Lichnerowicz operator \cite{Fursaev:1996uz}. The similar results were recently obtained in  \cite{He:2014gva}   using a somewhat simpler method   that can be further 
extended to compute the entropy in string theory.

The non-analytic term in the heat kernel for fields of spin $3/2$ is proportional to $\alpha$ and thus does not 
produce any contribution to the entropy. Therefore, we shall discuss only the case of the Lichnerowicz operator for a  field  of spin $s=2$.

Let us briefly outline the way this term emerges in the analysis of \cite{Fursaev:1996uz}. They consider the two dimensional case as the basic one and use the heat kernel obtained in two dimensions to derive the coefficients in higher dimensions by considering the products of two-dimensional space and $(d-2)$-dimensional surface $\Sigma$.
This approach is sufficient if we are interested only in coefficient $a_1$. Thus, they consider a singular two-dimensional sphere $S^2_\alpha$ which is obtained from the regular sphere with coordinates $(\theta,\tau)$ by identifying angle $\tau$ with period $2\pi\alpha$. They specifically consider the case when $\alpha<1$ that corresponds to an orbifold.
On such a singular sphere Fursaev and Miele construct two scalar functions 
\be
\phi^\pm_\alpha=\sin^\alpha \theta e^{\pm i \alpha\tau}
\lb{ff}
\ee
 that are eigen-functions of the Laplace operator
\be
-\nabla^2\phi^\pm_\alpha=\frac{(1+\alpha)}{\alpha^2}\phi^\pm_\alpha\, .
\lb{ef}
\ee
From these two functions one may construct 4 vectors $(V^\pm_\alpha)_\mu=\epsilon_\mu^{\ \nu}\partial _\nu \phi^\pm_\alpha$ and $(\bar{V}^\pm_\alpha)_\mu=\epsilon_{\mu\nu}(V^\pm_\alpha)^\nu$. For regular sphere $(\alpha=1)$ 2 vectors $V^\pm$ are the Killing vectors and the other two vectors
$\bar{V}^\pm$ are the conformal Killing vectors. Therefore, they can not be used to produce any physical gravitational perturbation.
On the other hand, when $\alpha\neq 1$ these 4 vectors are no more the Killing vectors and thus can be used to construct 4 gravitational modes 
\be
(h^\pm_\alpha)_{\mu\nu}=\nabla_\mu (V^\pm_\alpha)_\nu+\nabla_\nu (V^\pm_\alpha)_\mu\, , \,\, (\bar{h}^\pm_\alpha)_{\mu\nu}=\nabla_\mu (\bar{V}^\pm_\alpha)_\nu+\nabla_\nu (\bar{V}^\pm_\alpha)_\mu-g_{\mu\nu}\nabla_\alpha (\bar{V}^\pm_\alpha)^\alpha\, .
\lb{hV}
\ee
These modes are new compared to those gravitational modes which already existed for regular sphere $(\alpha=1)$. Each mode contributes $2\pi$ to the heat kernel coefficient 
$a_1$. For a higher dimensional space which is product of two dimensional space and a $(d-2)$-dimensional surface $\Sigma$ this results to modification 
\be
a_1^\Sigma=a_{1,L}^\Sigma+8\pi \int_\Sigma 1\ ,
\lb{aL}
\ee
where $a_{1,L}^\Sigma$ is the surface coefficient (\ref{con8}). We see that this new term does not depend on $\alpha$ and does not disappear when $\alpha=1$.

It is an important question whether these new modes (\ref{hV}) are normalizable.  One finds  for these modes that $h_{\theta\theta}\sim (\sin\theta)^{\frac{2}{\alpha}-4}$
so that they are normalizable with measure $\sin\theta d\theta$ only if $\alpha<1$. The latter condition is satisfied when $S^2_\alpha$ is orbifold.
On the other hand, these modes are non-normalizable if $\alpha>1$ and in particular they should be excluded when $\alpha=n$ is an integer.
Thus, for an $n$-fold cover of the two-dimensional sphere the gravitational modes are the same as for the regular sphere. The authors of \cite{Fursaev:1996uz} have speculated that
for $\alpha>1$ there may exist a self-adjoint extension of the Lichnerowicz operator for which the heat kernel is analytically continued to all values of $\alpha$.
However, it is more reasonable to think that, once we are dealing with some non-analyticity (between $\alpha<1$ and $\alpha=0$ cases) anyway, the case of $\alpha>1$ is analytically related to the regular case $\alpha=1$.
This is achieved by excluding the modes (\ref{hV}) from the spectrum that results in the same heat kernel (\ref{con8}) as we had before, i.e.  without the non-analytic term $8\pi$.
This consideration demonstrates that the orbifold and an $n$-fold cover are not in general analytically related to each other. In the context of the entropy calculation that uses the replica trick we are
dealing with an $n$-fold cover. We conclude that for the entropy calculation the non-analytic terms should not appear. The absence of these terms  is actually important for the renormalization of the entropy, both entanglement and conical. If the non-analytical terms were present neither of these entropies would be renormalized in the same way as the Newton constant.


\section{Can the Bekenstein-Hawking entropy be equal to  entanglement entropy?}
\setcounter{equation}0

Let us first consider the case when the only fields present are the matter fields, scalars or fermions, that couple minimally to gravity. For simplicity we shall consider the massless fields. The generalization to the massive field is straightforward.
Then, as we have already noted earlier,  the entanglement entropy and the Newton constant are renormalized in exactly  same way so that, provided there is no bare
Newton constant, i.e. the entire gravitational coupling $G_{ren}$ is induced, we have that
\be
{1\o 4G_{ren}}={1\o (4\pi)^{d-2 \o 2}(d-2)}  {1\o \epsilon^{d-2}}\sum_s N_s{{\cal D}_{s}(d)\o 6}\, ,
\lb{7.1}
\ee
where $N_s$ is the number of fields of spin $s$.  Here $\epsilon$ is the UV cut-off. By relation (\ref{7.1}) it is fixed to be proportional to the 
Planck length. If there is a black hole with horizon $\Sigma$ in the induced gravitational theory then it has the Bekenstein-Hawking entropy defined in terms of the gravitational coupling
constant $G_{ren}$ (\ref{7.1}) 
\be
S_{BH}=\frac{A(\Sigma)}{4G_{ren}}\, ,
\lb{7.2}
\ee
where $A(\Sigma)$ is the area of $\Sigma$. On the other hand, the entanglement entropy of $\Sigma$ due to the same set of fields is exactly the Bekenstein-Hawking entropy
\be
S_{ent}=\frac{A(\Sigma)}{4G_{ren}}=S_{BH}\, .
\lb{7.3}
\ee
This is guaranteed by the fact the Newton constant and the entanglement entropy are renormalized in exactly same way. Thus, if there were only matters fields, scalars and fermions, in the Universe then the above mechanism would be the natural explanation for the statistical origin of the black hole entropy.

However, in the observable Universe there exist not only matter fields but also the fields which mediate the interactions, gauge vector bosons and gravitons.
For these fields the renormalization of the Newton constant and the entanglement entropy is different due to the presence of the contact terms $c^{(s)}(d)$
that contribute negatively to the Newton constant and do not appear at all in the entanglement entropy. The presence of these fields in  our Universe 
makes impossible this simple identification of Bekenstein-Hawking entropy and the entanglement entropy that was just outlined. At this point it is tempting to
suggest (as  for instance in \cite{L}) that instead of the entanglement entropy we should consider the conical entropy   which, as we discussed this above, repeats identically the renormalization
of the Newton  constant. Therefore, extending the identification of the observable Newton constant (\ref{7.1}) to other spins, $s=1$ and $s=2$,
we would still have the equality between the Bekenstein-Hawking entropy and the conical entropy,
\be
S_{BH}=S_{con}\, .
\lb{7.4}
\ee
The only problem with this idea, as we already mentioned this earlier,  is that the conical entropy is not a priori positive definite (the positivity would impose extra constraint on the multiplet of fields) so that, most importantly, it is not determined by a well-defined density matrix and hence the desired statistical meaning of the entropy is lacking. 

In general, when the bare Newton constant $G$ is present, the renormalized gravitational coupling is defined by the relation
\be
{1\o 4G_{ren}}={1\o 4G}+{1\o (4\pi)^{d-2 \o 2}(d-2)}  {1\o \epsilon^{d-2}}\sum_s N_s({{\cal D}_{s}(d)\o 6}-c^{(s)}(d))\, .
\lb{7.5}
\ee
Notice that in the contact terms we include also the contribution of scalar particles, $c^{(s=0)}=\xi$, due to the phenomenological possibility \cite{Higgs} that  the Higgs boson may have a  negative (and even large $|\xi |\gg 1$)
non-minimal coupling  to gravity.

The question we want now to ask is whether it is possible to modify the above mechanism and have at the same time two things:

\noindent{\it 1.)} Gravitational constant is defined according to (\ref{7.5}) and, respectively, the Bekenstein-Hawking entropy is defined as 
$S_{BH}=A(\Sigma)/4G_{ren}$ in terms of this constant.

\noindent{\it 2.)} Entanglement entropy defined as
\be
S_{ent}={1\o (4\pi)^{d-2 \o 2}(d-2)}  {1\o \epsilon^{d-2}}\sum_s N_s{{\cal D}_{s}(d)\o 6}
\lb{7.6}
\ee
is identical to the Bekenstein-Hawking entropy,
\be
S_{BH}=S_{ent}\, .
\ee
The answer to this question is affirmative provided the following consistency condition 
\be
{1\o 4G}={1\o (4\pi)^{d-2 \o 2}(d-2)}  {1\o \epsilon^{d-2}}\sum_s N_s c^{(s)}(d)\, 
\lb{7.7}
\ee
is satisfied.
Since the bare Newton constant $G$ is a free parameter in the theory it always can be chosen to satisfy the condition (\ref{7.7}).
This condition simply means that the bare Newton constant $G$  cancels completely the contribution due to the contact terms in (\ref{7.5}) so that the renormalized 
Newton constant is defined only by the physical degrees of freedom
\be
{1\o 4G_{ren}}={1\o (4\pi)^{d-2 \o 2}(d-2)}  {1\o \epsilon^{d-2}}\sum_s N_s{{\cal D}_{s}(d)\o 6}
\lb{7.8}
\ee
in the same way as in (\ref{7.1}). The ratio of the bare and the renormalized  constants is determined only by the field multiplet
\be
\frac{G_{ren}}{G}=6\, \frac{\sum_s N_s c^{(s)}(d)}{\sum_s N_s {\cal D}_s(d)}\, .
\lb{7.9}
\ee
We see that the presence of the bare Newton constant $G$ is important for this mechanism to work. In the complete Bekenstein-Hawking entropy
there are two pieces which do not have a statistical meaning: the one due to the bare constant $G$ and the other due to the contact terms.
With the condition (\ref{7.7}) these two pieces mutually cancel and what is left has a perfect statistical meaning as entropy of entanglement.
As we have already mentioned above, the generalization of this mechanism and, in particular, of relations (\ref{7.7})-(\ref{7.9}) to the case of massive fields is straightforward and does not bring
any new principle issues.

A nice  bonus of the proposed mechanism is that both the entropy (\ref{7.6}) and the observable Newton constant  (\ref{7.8}) are determined only by the physical degrees of
freedom and, thus, do not depend on the gauge.

\section{Conclusions}
\setcounter{equation}0

As we show in this paper the analysis of the Newton constant and the entanglement entropy can be done in terms of the fields which appear in the low energy description.
The whole picture thus does not seem to depend on the details of the fundamental theory. In fact, this was anticipated by many authors who noticed  the 
universality of the Bekenstein-Hawking entropy which does not seem to depend on how the low energy gravitational physics is completed in the higher energy regime.
In the present context, however, the fundamental theory should give us a concrete realization of the mechanism proposed in this paper.
In particular, we  expect that the low energy parameters such as the bare Newton constant $G$  will arise unambiguously in the fundamental theory   
and hence the relation (\ref{7.7}) should be reproduced in a concrete calculation. 

The other lesson which we should learn from the above consideration is that the orbifold and the $n$-fold cover constructions are not always analytically related.
Even though the orbifolds look easier to make calculations on a better understanding of the $n$-fold covers is probably needed to make a further progress.

\bigskip
\section*{Acknowledgements} The most of this work was done when the author was visiting the Theory Division at CERN in June 2010
and the KITP  at UC Santa Barbara in May 2012. The author is grateful to  T. Damour, G. Dvali, M. Duff, T. Jacobson, W. Donnelly, A. Tseytlin  and A. Wall for useful discussions
and remarks.

\end{document}